FERMILAB-CONF-12-552-AD# FNAL PROTON SOURCE HIGH INTENSITY OPERATIONS AND BEAM LOSS CONTROL*

F.G. Garcia[#], W. Pellico, FNAL, Batavia, IL 60510, USA

## Abstract

The 40-year-old Fermilab Proton Source machines, constituted by the Pre-Injector, Linac and the synchrotron Booster, have been the workhorse of the Fermi National Accelerator Laboratory (Fermilab). During this time, the High Energy Physics Program has demanded an increase in proton throughput, especially during the past decade with the beginning of the neutrino program at Fermilab. In order to achieve a successful program, major upgrades and changes were made in Booster. Once again, the Proton Source has been charged to double their beam throughput, while maintain the present residual activation levels, to meet the laboratory Intensity Frontier program goals until new machines are built and operational to replace the Proton Source machines. This paper discusses the present performance of Booster and the plans involved in reaching even higher intensities.

## INTRODUCTION

The Fermilab Booster [1] is rapid cycling synchrotron which accelerates protons from the injection energy of 400 MeV to 8 GeV in 33 msec, at up to 15 Hz. It is 472 m in circumference and has a harmonic number of 84, with 96 combined function magnets distributed on a FDooDFo 24 symmetric lattice period. The Booster frequency changes rapidly through the accelerator cycle, from 37.9 – 52.8 MHz and there are 19 RF cavities in the machine. Early in 2000's the demand for protons increased 12-fold in comparison to the previous 10 years of Booster operations with the beginning of the neutrino program at Fermilab.

Present protons per batch in Booster are 4.5E12 at 7.5 Hz with 90% efficiency and 85% uptime. In the future, the required number of protons to the next generation of neutrino experiments has once again challenged the Proton Source: a factor of 2 more protons from the present running conditions is expected out of Booster. In order to achieve this demand, the number of cycles with beam will be increased rather than the intensity per cycle. Currently the Booster RF cavities do not possess appropriate cooling to run reliably at high repetition rate. Therefore, improvements in both hardware and operational efficiency of the Booster are required in order to have a successful physics program.

Therefore, the Proton Improvement Plan (PIP) [2] was established in 2010. The goal is to deliver 2.25 E17 protons per hour at 15 Hz by 2016. This increase in proton throughput has to be achieved by:

*Work supported under DOE contract DE-AC02-76CH03000
#fgarcia@fnal.gov

- maintaining 85% or higher availability;
- maintaining the same residual activation in the accelerator components.

## PRESENT BOOSTER PERFORMANCE

Figure 1 shows the protons delivered per day and the integrated protons delivered since 1992 up to May 2012.

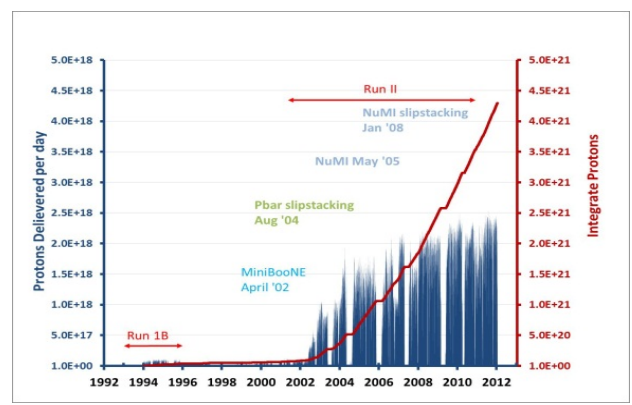

Figure 1: Booster integrated and per day protons delivery for the past 2 decades.

The primary users of the protons were the 8 GeV neutrino experiments at the Booster Neutrino Beam (BNB), whose repetition rate depends on other demands for protons, but typically run at 2 Hz up to 5 Hz. Another major consumer of protons is the 120 GeV neutrino experiments from the Neutrino at the Main Injector (NuMI). In this case, Booster provides 9 batches of 4.5E12 protons per pulse to Main Injector where the batches are slipstacked to generate 300 kW of beam power to the NuMI target.

The major projects that permitted Booster to run at higher intensities are quickly described in the following sections.

### New Corrector System

While the main lattice elements of the Booster ramp, the corrector system had historically operated DC. This means that the beam position moved on the order of several millimeters over the acceleration cycle. A multi pole magnet [3] was installed during shutdown periods in 2007 and completed in 2009 in the 48 locations around the ring with each capable of ramping which stabilized the Booster orbit through the accelerator cycle. Some major improvements in beam orbit control were obtained, such as better tune control at high energy, smaller emittance and better orbit coupling correction at high energy.

## Collimation System

In 2004 a new two-stage collimation system was implemented in Booster at period L5, L6 and L7 to aid deposition of uncontrolled losses at a location that is well shielded protecting essential machine elements for instance the RF system. The installation was complete in 2004 and after commissioning a significant reduction in activation in the RF stations were observed as shown in Figure 2.

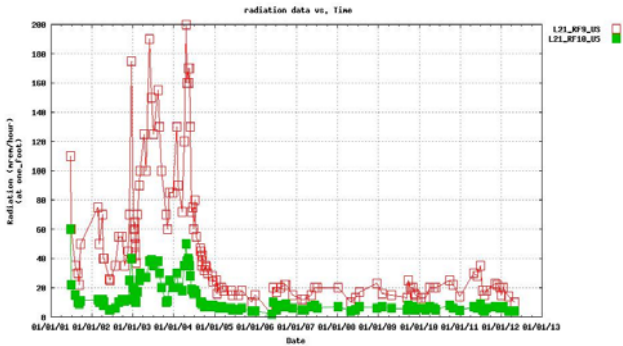

Figure 2: Example of radiation data at two Booster RF stations. A noticeable reduction in radiation is seen after 2004 when the collimators were installed.

## Beam Loss Control

Better beam loss control in the tunnel was implemented by adding an array of 50 interlock loss monitors arranged around the ring. Each loss monitor was set to a limit which roughly speaking was twice the levels at that location prior the high intensity running period. Later on the limits were fine tuned to the observed losses at a particular region.

In addition to the loss monitors, the average beam power loss is calculated by measuring the number of protons in the ring, weighted by the beam energy and integrating it through the cycle. Presently the limit is set to 525 W.

## Notch System

Booster beam has an abort gap created with the intent to reduce losses at the extraction region. 3% of beam at the injection energy is removed from the buckets to create the notch. By creating the notch at injection, the total energy loss is reduced by a factor of 20. This system was implemented for start of Run II and MiniBooNE. With the start of NuMI beam line experiments, in order to implement multi-batch scheme in Main Injector, the cogging system [4] was brought into operation. Due to the synchronization that is required between Booster and Main Injector, the notch for cogging events is created about 5 msec after injection. In both cases, the kicked beam is aimed to the collimator system. However, about 87% of the lost beam is deposited at the magnet pole tips. Booster has never lost a gradient magnet, but radiation is noticeable at this location.

The Fermilab Booster had made impressive progress in beam quality, designed, implemented and successfully commissioned upgrades and beam controls and monitoring systems in order to meet the goals for the MiniBooNE and NUMI/MINOS experiment. Despite the fact that Booster increased the beam throughput by 12-fold with only a factor of 3 in increase the total power loss in the tunnel is quite impressive. With all these upgrades, Booster operational beam power lies at 40 kW. Now Booster faces another challenge after the yearlong 2012 shutdown. The driving force for this long shutdown is to make the necessary Accelerator and NuMI Upgrades (ANU) for the NuMI Off-axis neutrino Appearance (NOvA) experiment. This experiment requires 50kW of beam power alone from Booster which is 25% more beam power than Booster has ever delivered to the entire laboratory physics program for the past decade. The PIP will address the technical aspect that is required in order to increase the repetition rate and maintain good availability and acceptable residual activation.

## ACHIEVING HIGH FLUX IN BOOSTER

As mentioned above, the next generation of neutrino experiment requires high intensity protons on target to successfully fulfil the physics case. In the case of NOvA it is expected up to 700 kW of peak proton power with $4.3E12$ protons per batch from Booster delivered to Recycler in 12 spills at 9 Hz to be injected in one extraction to Main Injector for accelerating up to 120 GeV in 1.33 sec. This corresponds to about 50 kW of proton power delivered by Booster to this experiment alone. To simultaneously support the BNB physics program and the high energy neutrino program it will require to run the Booster at 15 Hz. Figure 3 shows the proton delivery projection for the next decade.

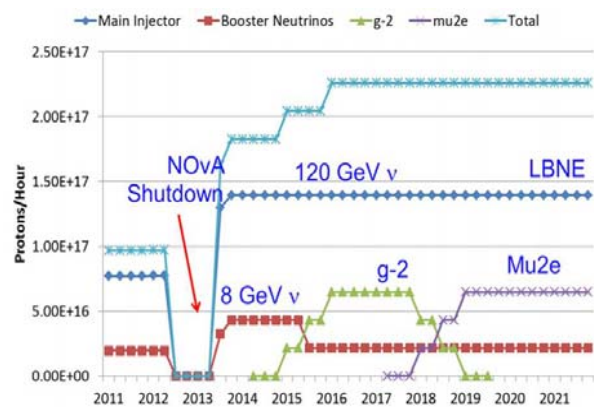

Figure 3: PIP protons economics for the next decade.

Therefore Booster is expected to deliver $2.25E17$ protons per hour at 15Hz by 2016. This requirement will necessitate a substantial upgrade of the RF power system in the Booster, in addition to improvements in the control of beam losses, shielding of the accelerator enclosure and improvement in reliability.

### RF Power System

Of the various contributors to limited repetition rates for the Booster accelerator, the Booster RF system has often been cited as a primary factor. The reason for that is that the system has never been designed to accelerate beam at the sustained rates now being expected. The fundamental limitations have been due to RF equipment heating. In order to improve the Booster RF reliability, the upgrade of the power amplifier is one of the main efforts inside PIP.

**Cavity and tuner refurbishment** The cavity and associated tuners, 3 per cavity, require cooling improvements to support higher repetition rates. Each cavity has to be removed from operation to be inspected, cleaned, tuners re-worked, vacuum certified and tested at 15 Hz to certify operations prior to being installed in the tunnel again.

The ferrite tuner cone cooling path had been disconnected many years ago because of water leak history. As part of the refurbishment each tuner needs to be completely disassembled and the cooling channels reworked. It is expected that some of the tuners cannot be refurbished and replacements will be needed. PIP has a task that will address the procurement and testing of new tuners.

**Solid State upgrade** The Booster RF system, among its kindred Main Injector RF systems, has the oldest equipment and exhibits, not surprisingly, the least reliability. The DA tubes and the Cascode sections of the cavity mounted Power Amplifier (PA) is especially vulnerable to more frequent failures. The repair of the RF system is also compromised by the increased activation of components.

It is expected that the greatest RF system reliability will come with the complete installation of a solid state RF driver and new Modulator in the equipment galleries and a new final stage amplifier at the enclosure cavity.

### Alignment and Aperture

Improving Booster acceptance will decrease operational beam losses. Aperture scans have been performed prior the shutdown in order to validate the programs and simulation model. One magnet move was performed based on the predictions suggested by the simulation based on the latest Booster alignment data. Figure 4 shows the results of the first magnet move performed prior the shutdown 2012. Improvement in the beam aperture at this location was achieved.

The approach is to proceed with the locations that make the largest contributions to the beam orbit imperfections.

### Notch and Cogging Upgrades

The Booster notcher system is used to create a gap in the beam for the extraction kickers. The phase I for this upgrade is to move the notcher kickers from Long 5 to Long 12 and install a new absorber. At 15 Hz the total beam power lost predicted in the tunnel is 270 W.

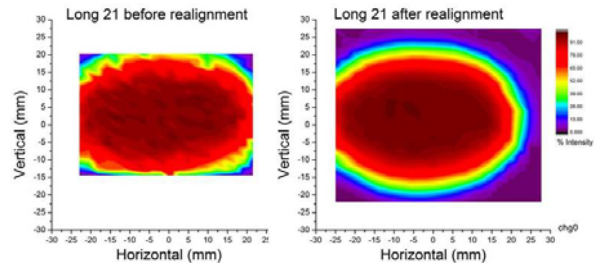

Figure 4: Aperture improvement by using lattice model combined with latest "as-found" metrology data.

As a phase II of this upgrade, a new complement of 6 short kickers with respective new power supplies will replace the existing system. This is expected to reduce the operational losses created by this system. Further efforts have been pursued such as remove the notch completely outside the Booster tunnel. If successful implemented, this system will be used as a secondary resource to clean the abort gap even further.

Finally, but not least, the cogging system will be improved in order to reduce the manipulation of the radial orbit. The new scheme will utilize the horizontal dipole correctors located at the maximum horizontal beta region to change the magnetic field during the acceleration cycle. Improved orbit stability enables better control of optics, tune and a more feasible collimation scheme to be implemented through all the accelerating cycle reducing beam losses.

### Radiation Shielding

Radiation damage to Booster components will remain an ongoing concern especially at expectations of higher proton throughput. During the revision of the Booster Shielding Assessment the biggest concern evolved around the 182 single-leg cable penetrations current filled with 12 ft of polyethylene beads. In a short summary, the number of active shielding controls will increase by 5 additional locations and work continues to identify the solution for the penetrations.

## CONCLUSIONS

The 40-year old Fermilab Booster has made tremendous progress toward meeting the demand of the Fermilab physics program. Another factor of 2 in protons deliverable is expected from Booster to support the future physics program at the laboratory. The Proton Source has been the workhorse of the Fermilab physics program and is it expected to be operational until 2025 when the next machine is supposed to be operational. Within PIP a number of improvements have been planned and there is optimism that Proton Source will be able to reach these goals.

## ACKNOWLEDGEMENTS

This work would not have been possible without the dedication and tireless efforts of many members of the Proton Source Department, Accelerator Division Support Departments, Accelerator Physics Center (APC), Technical Division (TD) and Particle Physics Division (PPD). The author is grateful for helpful discussions with Alexandr Drozhdin, Salah Chaurize, Duane Newhart, William Pellico, Kiyomi Seiya, Todd Sullivan.

## REFERENCES

[1] E. L. Hubbard *et al.*, "Booster Synchrotron", Fermilab-TM-405 (1973).
[2] F.G. Garcia *et al.,* "PIP Design Handbook", Fermilab-Beams-doc-4053-v4, (2012).
[3] E.J. Prebys *et al.*, "Booster Corrector System Specification", Fermilab Beams-doc-1881-v5, (2006).
[4] R.Zwaska *et al.,* "Cycle-to-Cycle Extraction Synchronization of the Fermilab Booster for Multiple Batch Injection to the Main Injector", PAC2005, Knoxville, Tennessee, USA, 2005, p1802.